\documentclass{article}
\usepackage{ijcai18}

\usepackage{times}
\usepackage[table]{xcolor}
\usepackage{soul}
\usepackage[utf8]{inputenc}
\usepackage[small]{caption}
\usepackage{gensymb}

\usepackage{graphicx}
\usepackage{url}
\usepackage{color, colortbl}
\title{A Computational Framework for Modelling and Analyzing Ice
  Storms}

\author{\\
  \{\textbf{Keywords:} Computational Sustainability and Modelling, Machine Learning\}\\
  Ranjini Swaminathan$^1$,
  Mohan Sridharan$^2$,
  Katharine Hayhoe$^1$
  \\
  ranjinis@gmail.com,
  m.sridharan@auckland.ac.nz,
  katharine.hayhoe@ttu.edu
\\
$^1$ Texas Tech University,
$^2$ The University of Auckland\\
}
\begin{document}

\maketitle

\begin{abstract}
  Ice storms are extreme weather events that can have devastating
  implications for the sustainability of natural ecosystems as well as
  man made infrastructure. Ice storms are caused by a complex mix of
  atmospheric conditions and are among the least understood of severe
  weather events. Our ability to model ice storms and characterize
  storm features will go a long way towards both enabling support
  systems that offset storm impacts and increasing our understanding
  of ice storms. In this paper, we present a holistic computational
  framework to answer key questions of interest about ice storms. We
  model ice storms as a function of relevant surface and atmospheric
  variables. We learn these models by adapting and applying supervised
  and unsupervised machine learning algorithms on data with missing or
  incorrect labels. We also include a knowledge representation module
  that reasons with domain knowledge to revise the output of the
  learned models. Our models are trained using reanalysis data and
  historical records of storm events. We evaluate these models on
  reanalyis data as well as Global Climate Model (GCM) data for
  historical and future climate change scenarios. Furthermore, we
  discuss the use of appropriate bias correction approaches to run
  such modeling frameworks with GCM data.
\end{abstract}

\section{Introduction}
Ice storms are extreme weather events that can cause extensive and
permanent devastation to ecosystems, infrastructure and life. They are
characterized by freezing rain causing ice to glaze over exposed
surfaces such as roads, power lines and tree branches. They are not
necessarily high precipitation (rainfall) events; even small amounts
of ice accumulation can increase the branch weight of trees by up to
one hundred times its actual weight. The same applies to power lines
and even larger communication towers. Ice deposited on tree branches
can lead to severe destruction of local ecosystems, and fallen tree
branches can damage life and property, as well as obstruct essential
pathways and roads. Severed power lines have been
responsible for extended power outages, with significant damage to
infrastructure and in some cases even resulting in human
casualties~\cite{jones_report_98}. Accidents due to icy road
conditions also contribute to property loss and the loss of human
lives. Losses from the $1998$ ice storm that affected north eastern USA and south
eastern Canada were estimated at 6.2 billion U.S dollars with less
than one half of the amount insured. More than four million people
were left without power and about $40$ people lost their lives due to
icy roads and lack of essential services. Combined losses due to ice
storms in recent years amount to billions of dollars in measurable
losses~\cite{gyakum_mwr_01,irland_jf_98}. To plan and
prepare for ice storm impacts, stakeholders need to understand how the
prevalence, duration and intensity of storms may change in the
future, which proves challenging for several reasons.  First, there is
insufficient scientific understanding of anything other than the basic
physical atmospheric processes that cause ice
storms~\cite{kunkel-bams:13}.  We can study storms in the past to
improve our understanding about storms, but historical storm record
accuracy and availability vary according to factors such as population
density and infrastructure distribution~\cite{cutter_nas_08}. For
instance, ice accumulation information in sparsely populated or
heavily forested areas may not be recorded with as much rigour as
metropolitan areas. Furthermore, databases that record storms and
other extreme events typically suffer from reporting biases, often
resulting in incomplete and disparate information~\cite{gall_bams_09}.
Finally, any information learned from past records will have to be
considered in the context of future change to provide similar
information about ice storms in the future.

\subsection{Problem Definition}
In this paper, we propose a holistic framework to address the
challenges arising from inadequate ice storm data and domain
expertise, and to provide relevant information about ice storm
occurrences in the future.  We use historical records of ice storms
from storm databases and information about atmospheric variables
(e.g., temperature, humidity etc.) on storm and non-storm days from
gridded reanalysis data. Reanalysis is an approximation of observed
weather data generated by a weather model constrained by observations,
which we use to learn models that explain ice storm behaviour. We then
use Global Climate Model (GCM) simulations to evaluate our learned
models and develop ice storm projections under future scenarios. Our
framework addresses relevant aspects of ice storms and how they may
change in the future as follows:
\begin{itemize}
\item \textbf{Ice storm prevalence, including frequency of
    occurrence:} we build a storm detector using historical storm data
  records and large scale, relatively coarse surface and upper
  atmospheric conditions accompanying the historical storms.

\item \textbf{Ice storm intensity:} we characterize storm features
  such as intensity by matching salient features of objectively
  identified storms (in simulated GCM data) with those of historical
  storms using an hierarchical agglomerative clustering algorithm.

\item \textbf{Ice storm information for the future:} we develop
  appropriate bias correction approaches to make any models learned on
  historical reanalysis data compatible with simulated GCM projections
  for the future.

\item \textbf{Domain knowledge representation:} we develop a module
  that facilitates the incorporation of newly acquired domain
  knowledge to further refine our modelling outputs.
\end{itemize}
While we provide experimental results for a specific natural
phenomenon (\textit{icestorms}) occurring in a particular region
\textit{northeastern continental USA} in this paper, our framework can
be adapted for similar extreme weather events and different geographic
locations. Furthermore, we illustrate the interdisciplinary nature of
our work and show how machine learning and AI methods can be adapted
to formulate and address a key problem in climate science and
resilience planning.

\section{Related Work}
\label{sec:relatedwork}
A study of the temporal and spatial distribution of freezing rain events associated with ice storms in the contiguous USA show 
that about half these events occur in the northeastern region~\cite{changnon_jam_03}. The national maximums were found to be in New York and 
Pennsylvania, as a result of storm-favorable weather conditions.~\cite{rauber_jam_01}  study typical 
atmospheric patterns, including those associated with topography, that cause most storms in the north and southeastern regions of the USA. Their 
key findings include the fact that that the vertical temperature profile and surface and upper air wind directions were characteristic of the overall 
archetypical storm patterns.~\cite{castellano_thesis_12} studied the atmospheric conditions associated with northeastern ice storms, including synoptic 
scale movement of moisture and temperature. These analyses are all grounded in observational data at relatively high spatial and temporal resolutions. 
Currently, ice storm forecasts are calculated by feeding this data into computationally intensive weather models that produce composite maps which 
are then subjectively analyzed by domain experts. 
The necessity for  both high resolution data and human subject experts is further compounded by the question of how these ice storms will be 
affected by changing climate patterns in the future. According to the Intergovernmental Panel of Climate Change's recent report, milder winter 
temperatures in the future could cause an increase in freezing rain, especially if average daily temperatures fluctuate about the freezing point ~\cite{field_ipcc_12}. 
Global Climate Models (GCMs) use the laws of physics to simulate atmospheric circulation patterns, and are capable of generating three-dimensional 
projections for different atmospheric variables under varied future climate scenarios~\cite{taylor_bams_12}. GCM simulations generate these projections 
at coarser spatial (250km vs 32km)  and temporal (daily vs hourly) resolutions than observed data. \cite{cheng_ao_11} use a 
process called statistical downscaling to combine observations with GCM output to generate higher resolution projections, and apply the resulting output fields
to study changes in future occurrences of ice storms in Canada. More recently, advanced machine learning techniques such as deep networks 
have been used to understand extreme events such as tropical cyclones and weather fronts by studying large synoptic scale patterns  ~\cite{liu_arxiv_17} as well as 
for generating higher resolution data to study these extreme events ~\cite{vandal_kdd_17}.  As far as we are aware,~\cite{swaminathan_ausai_15} are the only 
others who attempt to objectively identify large synoptic scale patterns for ice storms using advanced machine learning algorithms. However, their choice of 
climate variables was incompatible with many GCMs and so cannot be used in ensemble model experiments for future scenarios. Furthermore, their experiments were 
restricted to historical reanalysis output, and do not discuss integrating domain knowledge or identifying other storm characteristics of interest. 
%Here, we expand
%on that initial work.

\section{Modeling Framework}
\begin{figure*}[tb]
  \begin{center}
    \includegraphics [width=0.9\textwidth]{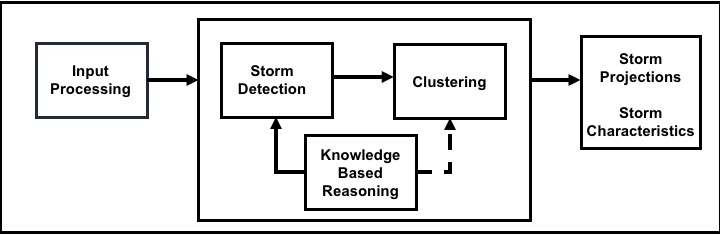}
    %\vspace{-1em}
    \caption{Overview of the ice storm modelling framework. The bold
      and dashed arrows indicate links that have been already been
      implemented and future work respectively.}
    \label{fig:overview}
  \end{center}
  \vspace{-1em}
\end{figure*}

We present an overview of our ice storm modelling framework in
Figure~\ref{fig:overview}. The framework includes an \emph{Input
  Preprocessing} stage which includes (at the minimum) extracting
dates from a storm data base, acquiring Global Climate Model and
reanalysis data, and regridding the data to appropriate resolutions.
The \emph{Storm Detection} and \emph{Clustering} modules are for
finding ice storms and understanding storm characteristics. The
\emph{Knowledge Based Reasoning} module is currently linked only to
the Storm Detection step but can potentially add information to the
Clustering process too. We get two main outputs: storm projections and
storm characteristics, which can be further analysed to get
information such as storm frequency, duration and seasonality.

\subsection{Ice Storm Detection} 
\label{subsec:stormdetect}
The first component of our ice storm modelling framework is a storm
detector that learns from historical storm data to recognize patterns
when presented with relevant atmospheric conditions. Our geographic
region of interest is the northeastern continental USA from eastern
Ohio across to Maine, and to Virginia in the south. To identify
synoptic or large scale patterns that may indicate ice storms, we look
at atmospheric variables between 55\degree N to 24\degree N and
50\degree W to 94\degree W for the winter months from October to
April.  We combined this information with historical records of storm
events from the National Climatic Data Center's Storm Data
database~\cite{ncdc_storm_db_18} to learn a storm detection model. We
also included storm events from the U.S Army Corps of Engineers'
Damaging Ice Storm GIS database that records storm footprints as
storms progress through different states in the
country~\cite{usac_storm_footprint_18}. A literature survey, as
described in Section~\ref{sec:relatedwork}, yielded domain expertise
that could then be applied to determine relevant near-surface and
upper atmospheric (at various pressure levels above the earth)
variables that could potentially indicate the presence of an ice storm
in the region. The features we use in our model are:
\textit{geopotential height or the gravity-adjusted height of pressure
  levels at 250mb, 500mb, 700mb and 850mb; temperature and specific
  humidity at the surface and at pressure levels 700mb and 850mb; and
  finally wind direction at the surface and at pressure level 850mb}.
We posed this as a classification problem where the chosen climate
variables in a 3-dimensional space are features that interact with
each other to either produce ice storm conditions (positive labels) or
not (negative labels). Our training dataset consisted of all winter
month days in the historical time period 1979-2008, which included
both negative days as well as positive days as identified by either
the Storm Data database or the Damaging Ice Storm database. Combining
these two datasets gave a total of \emph{493} individual days when
storms were recorded, with many of the days being part of multi-day
storm events; grouping together successive storm days, we find a total
of \emph{130} multi-day storm events in the thirty year period. It is
important to note that the observational data sources contain known
inconsistencies and inadequacies. It is possible that some storms were
not recorded if the impacts were not experienced in a well populated
zone. In the case of the storm footprint data, the days spanning a
storm event may include days where the storm was actually felt in
states outside the northeast USA. At this time, the available data
does not provide any way to prune out or add to the current list of
storms, but we consider these facts when we discuss our results.  For
the identified storm days, we would ideally use as feature values the
actual recorded quantities for the different climate variables.
However, since the balloon or radiosonde network that makes these
observations is very sparse, we instead use reanalysis data from the
NCEP North American Region Reanalysis (NARR) project. NARR was created
by taking all observed recordings during the period 1979-2008 and
assimilating them into high resolution numerical models to generate a
dynamically consistent climate state at each time
step~\cite{mesinger_bams_06}.  NARR output is available at 3-hourly
intervals each day and at a resolution of 32km, meaning that a single
reading would apply to a 32kmx32km grid cell at a certain height or
pressure level above the ground. With the number of variables and the
geographical area (about 145x147 grid cells) selected for this
analysis, this yielded over two million features. Since our models were
to be applied to GCM simulations which typically have much coarser
resolutions, we computed daily averages and smoothed the data with a
5x5 sub mask to arrive at $\sim$10,000 features.
 
Ice storms are relatively rare events as reflected in the positive to
negative sample ratio; even with using all the winter dates in the
selected time period, we had very few training samples relative to the
number of features, thereby rendering popular Neural
Network~\cite{gardner_ae_98} or Deep Learning~\cite{lecun_nature_15}
methods unsuitable for our task. We experimentally determined through
cross-validation experiments on the reanalysis data that a Support
Vector based classifier implementing a Sequential Minimal Optimization
algorithm~\cite{platt_book_98} and a polynomial kernel provided the
best classification results.

\subsection{Bias Correction for GCM Data}
\label{subsec:bc}
Once a storm classification model is learned using reanalysis data, we
could potentially use this model to analyze the frequency of such
storms in future climate scenarios. As mentioned earlier, Global
Climate Model (GCMs) are the primary sources of future climate
simulations. They generate consistent gridded output fields for
variables and at pressure levels similar to reanalysis, at somewhat
coarser spatial resolution, and at daily or sometimes even sub-daily
resolution for time periods generally ranging from 1900 to 2100 or
beyond. GCM simulations driven by future scenarios of human forcing
provide the basis for assessing the potential impacts of human-induced
climate change on a broad range of natural phenomena and geographic
regions. However, GCMs also exhibit biases or systemic errors in their
atmospheric circulation patterns compared to reanalysis, a result of
their lower spatial resolution, simplified physics and thermodynamic
processes, numerical schemes, and incomplete knowledge of physical
climate processes~\cite{navarro_ccafs_15}. Biases in GCMs relative to
observations can be significant, emphasizing the need to bias-correct
raw climate model outputs to better mimic observed patterns.
%\cite{maraun_cccr_16} reviews standard bias
%correction techniques in the climate sciences including the Delta and
%the Quantile Mapping methods. 
Traditionally, standard bias correction methods are applied to a
single variable such as temperature or precipitation. In our case, the
data is high dimensional.  Correcting along each dimension based on
observed data will not necessarily result in a bias correction in the
high-dimensional space. We therefore implemented bias correction with
bootstrap aggregation or bagging on the output obtained by applying
the storm detection model to the GCM output fields, rather than the
GCM output fields themselves, and obtained final classification labels
by consensus voting over all the bootstrap
subsets~\cite{friedman_slbook_01}.  For comparison purposes, we also
ran classification experiments on GCM data with standard bias
correction along each dimension. This was done by calculating a
six-week average around each individual winter date over a thirty year
climatological period for both reanalysis and GCM data. Bias was then
computed as $bias = model - reanalysis$ and subtracted from the model
variable data for each date and each variable.

\subsection{Determining Storm Characteristics with Hierarchical
  Agglomerative Clustering}
\label{subsec:hac}
One of the key pieces of information regarding ice storms that
stakeholders care about is storm intensity, or how much of an impact
an ice storm event will have. However, this measure tends to be very
subjective and frequently unreliable as it depends on presence of
infrastructure, type of ecosystem, and population distribution in the
areas affected by the storm.
%\cite{changnon_jam_03c} specifically
%discusses how the Storm Data database has incomplete information
%regarding losses and impacts measured. 
The Damaging Ice Storm Data set has text narratives describing the
effects of the storm in individual states and counties which makes for
a rich depository but requires extensive natural language processing
to extract this information in meaningful quantitative ways. We thus
decided to focus on the climatological aspects of ice storm events as
being a more objective measure of storm strength. We agree
with~\cite{rauber_jam_01} that archetypal storm categories exist, but
we do not necessarily know what they are at daily resolution and
synoptic scales. We thus frame this as an unsupervised learning
problem, where we neither have intensity (or category) labels for each
storm event nor have the number of such categories, but are aware that
they can be grouped. We cluster known storm features (using ground
truth data for recorded historical storm events) with objectively
detected storm day features (as obtained from classification
experiments). This way we ensure that storm intensity is a
scientifically measurable metric, independent of any specific impact
analysis, but one that can still be matched to impacts of previously
seen storms by looking at which reanalysis storm dates in the
historical databases and which positively classified dates are
together in each cluster. We use an hierarchical agglomerative
clustering algorithm to cluster reanalysis storm day features with GCM
projected features as described in~\cite{duda_pcbook_73}.  From the
dendrogram obtained from the clustering algorithm, we determine the
optimal number of clusters or categories for the storm events under
consideration. Since storm features encode temporal and spatial
information, we believe that clustering can give us further insights
into these aspects of storm events as well. This is by no means the
only approach to clustering or grouping storms, but we use this
algorithm to explore the extent to which clustering may improve our
understanding about the diversity of ice storm types and impacts.

\subsection{Domain Knowledge Representation}
\label{subsec:kr}
The overall goal of our modelling framework is to improve our
knowledge and understanding of ice storm events occurring in the
northeastern USA. However, as more research goes into studying ice
storms, including our own work with this framework, domain expertise
in the area continues to improve. We therefore introduced a module in
the framework that enables us to add growing domain knowledge, and
apply it to refine our outputs such as storm frequency and intensity
projections. We draw on well-established knowledge representation and
reasoning methods, and adapt them for our
application~\cite{brachman_krbook_92}. Our current framework
incorporates one such aspect of domain information as a pilot study.
Specifically, a visualization of composite maps for the variable
geopotential height at pressure level 500 mb and domain knowledge
about the influence of this geopotential height on storm conditions
was encoded as a rule to detect closed loop contours for this
particular variable on storm days. Such closed loop contours represent
low pressure systems that are considered to be a strong indicator of
storm conditions. In Section~\ref{sec:results}, we discuss the implications of this rule on
analyzing the false positives in our storm detection results to
illustrate how domain knowledge can have a significant role to play in
climate science domains.

\section{Experimental Results}
\label{sec:results}
\begin{table*}[tb]
\begin{tabular}{|p{50mm}|p{10mm}|p{10mm}|p{10mm}|p{10mm}|p{10mm}|p{10mm}|p{10mm}|p{15mm}|}

\hline
\textbf{Data Set} & \textbf{Oct} & \textbf{Nov} & \textbf{Dec} & \textbf{Jan} & \textbf{Feb} & \textbf{Mar} & \textbf{Apr} & \textbf{Total}\\
\hline
\rowcolor{blue!25}
Actual Storm Events &  1 & 6 &24  &39  & 28 &30  &2 & 130\\
\hline
%\rowcolor{red!25}
Raw Reanalysis Data & 3 & 14 &34&54&46&37&5&193 \\
\hline
\rowcolor{black!25}
Reanalysis (bootstrap) &1 &  6&21 &39  &27  &29 & 4&127 \\
\hline
Raw GCM Data &5 &15 &54 &80  &79 & 58&24  &315 \\
\hline
\rowcolor{black!25}
GCM Data (standard) &4 &  53& 61 &  102& 89&58 & 18&385 \\
\hline
\rowcolor{black!25}
GCM Data (bootstrap) &0  & 8 & 32 &57 & 53 &39  & 9 &198 \\
\hline

\end{tabular}
\caption{Storm events detected with the support vector based classification model for the period 1979-2008. Results are shown on GCM and reanalysis data with bias corrected data rows highlighted in grey. We show bias correction results using both standard and bootstrap methods with the GCM data.}
\label{table:smo-results-multiday-bootstrap-10}

\end{table*}

\begin{table*}[tb]
\begin{tabular}{|p{50mm}|p{10mm}|p{10mm}|p{10mm}|p{10mm}|p{10mm}|p{10mm}|p{10mm}|p{15mm}|}
\hline
\textbf{Data Set} & \textbf{Oct} & \textbf{Nov} & \textbf{Dec} & \textbf{Jan} & \textbf{Feb} & \textbf{Mar} & \textbf{Apr} & \textbf{Total}\\
\hline
\rowcolor{blue!25}
Actual Storm Events (historical) & 0.33  & 2 &  8  & 13   & 9.33 & 10  &0.67 &  43.33\\
\hline
GCM Data (historical) & 0  & 2.7 & 10.67  & 19 & 17.67  & 13  & 3  & 66 \\
\hline
GCM Data (2026-2035) & 1 & 1 & 8 & 15 &27 & 12 & 2 & 66 \\
\hline
GCM Data (2086-2095) &0 & 4 & 15 & 18 & 15 & 8 & 2 & 62 \\
\hline
\end{tabular}
\caption{Storm events detected with the support vector based classification model. Results are shown for historical GCM data averaged for a ten year period and two future ten year scenarios. All classification labels were bias corrected using bootstrap aggregation. }
\label{table:bc-10-gfdl-future}
\vspace{-1em}
\end{table*}

\begin{figure*}[tb]
  \begin{center}
    \includegraphics [width=0.9\textwidth,height=0.3\textheight]{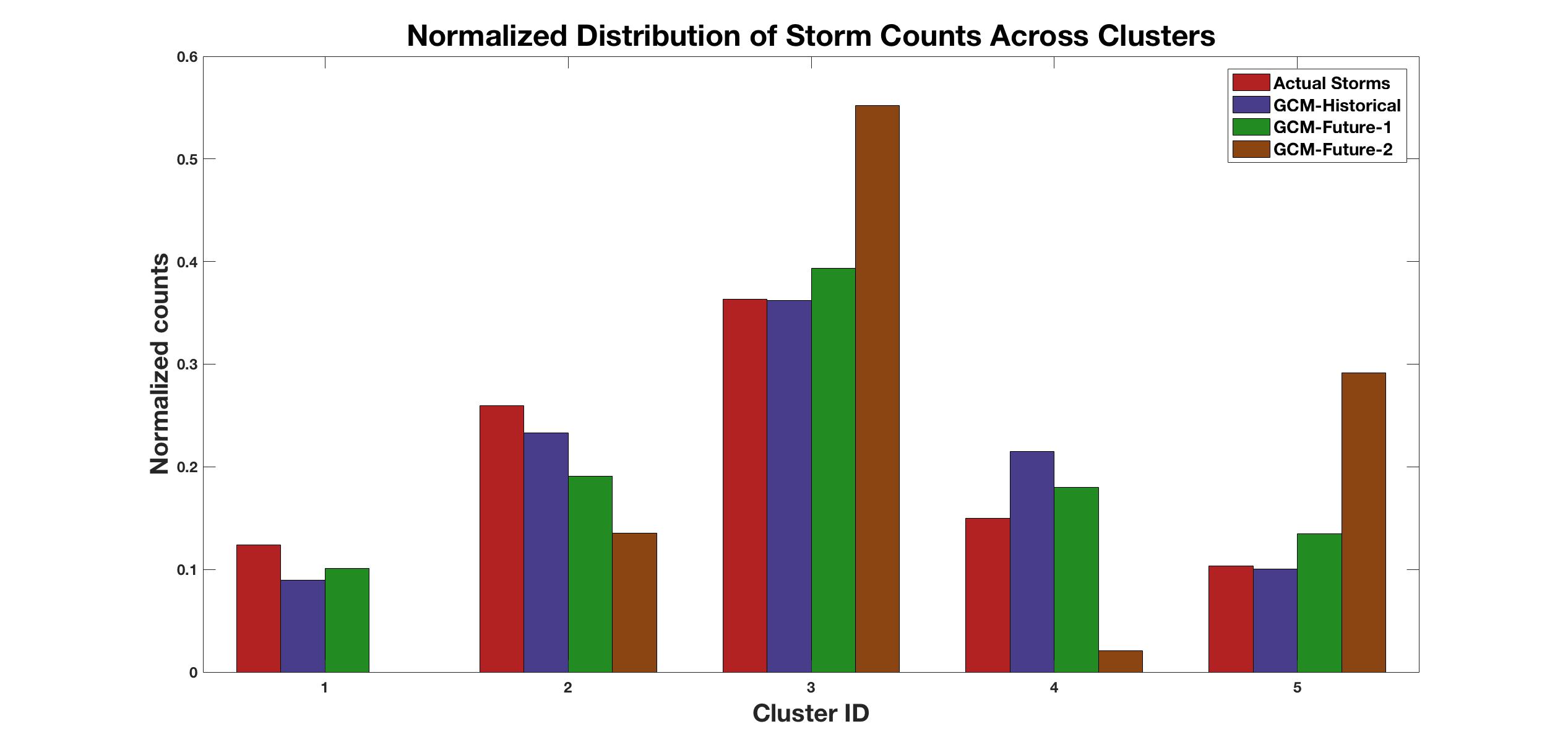}
    \vspace{-1em}
    \caption{Normalized counts of storms across clusters for reanalysis
      and GCM data including historical and future scenarios. GCM-Future-1 is the period (2026 - 2035) and GCM-Future-2 is the period (2086-2095).}
  \label{fig:norm_cluster_counts}
  \end{center}

  \vspace{-1em}
\end{figure*}

This section summarises the experimental results for each component of
our ice storm modelling framework. All GCM results are based on
simulations by the GFDL HIRAM high-resolution climate
model~\cite{dixon_cc_16} for the historical period between 1979 and
2008, and for two future time periods (2026-2035 and 2086-2095) under
a higher future scenario (RCP8.5) for which the relevant climate
variable data is available.

\textbf{Storm Detection:} Table
\ref{table:smo-results-multiday-bootstrap-10} summarizes our ability
to detect storms in the reanalysis data,
 %Table\ref{table:smo-results-multiday-bootstrap-10}  
% summarizes our ability™ to detect storms in the reanalysis data, 
raw GCM data and two kinds of bias corrected GCM data, and compares it
to the historical number of storms in the 30 year historical period
(which themselves contain known biases, as discussed previously).  The
two kinds of bias correction performed on the GCM data were the
standard and bootstrap method, as discussed in Section
\ref{subsec:bc}. Bootstrap consensus labelling results in simulated
historical storm count numbers very close to the actual storms
identified in the reanalysis output, providing a baseline for GCM
experiments. Note that standard bias correction does not apply to
reanalysis data since there is no bias there to correct.  It is also
important to note that the storm detection model recognizes more
storms in both historical reanalysis and GCM output than were
identified in the databases. We attribute this bias to the fact that
the ground truth data is known to be incomplete and subject to error
and reporting bias.

Interestingly, the standard bias-corrected GCM simulations provide
poor results. This indicates that traditional bias correction methods
do not work well with the kind of high dimensional data used here,
justifying the need for more advanced statistical methods.  Table
\ref{table:bc-10-gfdl-future} shows results on historical and future
GCM scenarios for a ten year period in comparison with the number of
actual storms typically observed in a historical ten year period. The
number of storm events in the future period are not significantly
different from those observed in the past. However, this time period
is too short to capture the uncertainty in natural variability and
arrive at any definitive conclusions regarding the impact of
human-induced warming on the frequency or temporal distribution of ice
storms over the northeast US. Instead, we propose to use an ensemble
of multiple GCMs with longer and continuous time series to increase
the sample size of the future projections \cite{tebaldi_royal_07}.

\textbf{Hierarchical Clustering:} Examining the clustering results in
Figure~\ref{fig:norm_cluster_counts} also reveals that the cluster
distribution of storms identified in the historical GCM simulation
closely matches that of the reanalysis data. This suggests, once
again, that the GCM is able to simulate a range of, or various types
of, events that resemble real-world ice storms to a sufficient extent
that the model trained on reanalysis is able to not only identify them
but also to group them into similar types of events. We analyzed the
cluster compositions in different ways such as the average length of
storms in each cluster and the day of storm in each cluster (e.g., did
all first days in storm events fall in a single cluster) but did not
see strong patterns. However, we noticed that starting from 1979 to
2008, the number of storms shifted across clusters with Cluster 3
seeing an increase in storms. In the two future scenarios, we again
notice that Cluster 3 has the highest membership, indicating that we
will likely see more storms like the ones in Cluster 3. We then looked
up some of the worst ice storms seen in the north eastern USA region
from our records collected for this project and an online search. We
found that in the list of worst storms to affect the region, almost
all (March 1991, January 1998, Dec 2007, April 2003, January 2007,
December 2007 and December 2008) were grouped in Cluster 3. We also
noted that all the dates that the storm lasted over the northeast USA
region were grouped under Cluster 3. In the case of storms with
footprints (storm forms in a different region and moves to the
northeast), we noticed a trend where the days when the storm was
likely outside the region of interest got assigned to a different
cluster. This leads us to conclude two things: (i) that we are able to
successfully use our hierarchical agglomerative clustering model to
capture storm intensity as a reflection of damages and impacts; and
(ii) that future storms are likely going to be more intense or fall in
the category of \emph{worst storms}. As in the case of storm detection
experiments, we propose to run ensemble runs with multiple GCMs to
increase the certainty of our findings. We also noticed that none of
the storm events occurring in the month of April were ever assigned to
Cluster 1. This could mean that April storms have a specific
characteristic that we have not yet been able to determine but we are
investigating ways to get domain expert feedback on these storms to
understand the significance of the cluster assignment.

\textbf{Domain Knowledge Representation:} Domain experts consider the
presence of a low pressure system as a strong, but by no means
exclusive, indicator of storm conditions. We conducted a simple
experiment where we applied this rule, and found that while the
majority of ground truth storm days did contain a closed low pressure
system, there were many days that did not (likely because they were
part of a multi-day storm footprint tracking event and the low
pressure system had either not moved in or already moved out of the
study area) and some days that did were not identified as ice storms
(because the criteria for ice storms includes very specific surface
conditions, notably a warmer layer of air overlying a cooler layer,
that are not present in every winter storm). We also found that our
storm detection system captured this hidden feature even though it was
not explicitly specified and some of the false positives in our result
were actually storm days with this low pressure system. By applying
this rule to the output of the storm detection module, we were able to
explain our so called \emph{false positives}. Adding more such rules
can help prune out or add confidence to our classification outputs. We
implemented this module as a pilot study and we see that it can be
applied as a pre-processing step to improve the quality of ground
truth data or as a post-processing step to further refine our
framework outputs.

\section{Contributions}
The ice storm modelling framework described in this paper addresses a
challenging and complex problem in environmental science that has
significant sustainability implications to our society. We answer key
questions regarding ice storms that are relevant to stakeholders who
undertake impact analysis projects. Climate and environmental sciences
offer novel and challenging problems in terms of data complexity,
uncertainty quantification at never seen before scales of data. Our
framework illustrates that successful solutions to such problems must
involve the application of interdisciplinary domain expertise and
carefully tailored solutions to different aspects of the whole problem
by posing the right questions instead of simply applying standard
machine learning approaches. This framework can also be readily
generalized to other geographic locations in the world and even to
understand similar synoptic scale weather events such as
hurricanes. We show that important features of such events like the
formation of a lower pressure system in ice storms can also be
captured without being explicitly modelled. Finally, we also show that
domain knowledge can be used in conjunction with statistical methods
to add value to research solutions in such complex domains.

\textbf{Threats to Validity} The process of modelling complex weather
phenomena and making future projections or predictions is made
challenging by many factors. First, ground truth data used for
training or learning models can be inaccurate, inadequate and
incomplete. Second, models learned with high resolution observed and
reanalysis data introduce a bias when applied on GCM data. Third, is
the issue of non-stationarity where the relationship between historic
climate model and observed variables will evolve over time and we do
not have the ability yet to forecast this effectively. Finally, though
GCMs are our primary source for future climate projections, they are
only approximations of the physical equations representative of
atmospheric processes and are further limited by computational
constraints.  Each stage in the process adds uncertainty and noise and
it is therefore important for us to be aware of them when using the
outputs of the framework.

\subsection{Future Work}
We believe that the work done so far on this project provides a broad based platform to pursue further research in several directions. First, we need to integrate an ensemble of GCMs with such frameworks to reduce climate modelling uncertainties and capture the range of possibilities under various possible future scenarios.  We would like to see if the knowledge representation module can be used to improve the quality of our ground truth so we can get better projections for the future. Our clustering results can be further explored to investigate different aspects of the storms such as geographic locations or length of storms which are all important factors to improve our understanding of ice storms. Finally, we would like to adapt this framework to study ice storms in other geographic regions and other similar weather phenomena.

\vspace{-10 pts}
\bibliographystyle{named}
%\bibliography{ijcai18}

\end{document}